       \newcommand{\beq}{\begin{equation}}
       \newcommand{\eeq}{\end{equation}}
       \newcommand{\beqa}{\begin{eqnarray}}
       \newcommand{\eeqa}{\end{eqnarray}}
       \newcommand{\beqas}{\begin{eqnarray*}}
       \newcommand{\eeqas}{\end{eqnarray*}}
       
       \newcommand{\bnab}{\mbox{\boldmath ${\nabla}$}}

       \newcommand{\x}{\mbox{\boldmath$\times$}}

       \newcommand{\bb}{{\mathbf b}}

       \newcommand{\be}{{\mathbf e}}

       \newcommand{\bv}{{\mathbf v}}

       \newcommand{\bA}{{\mathbf A}}
       \newcommand{\bB}{{\mathbf B}}

       \newcommand{\bE}{{\mathbf E}}

       \newcommand{\bR}{{\mathbf R}}

       \newcommand{\dadb}[2]{\frac{{  d}#1}{{  d}#2}}

       \newcommand{\paraparb}[2]{\frac{\partial #1}{\partial #2}}

\newcommand{\paral}{\parallel}

\documentclass[12pt]{article}
\usepackage[dvips]{graphicx}\usepackage[dvips]{graphics}
\usepackage{enumerate}
\usepackage{psfrag}
\usepackage{color}
\begin{document}
\begin{center}
{\large \bf
 A comparison of Vlasov with drift kinetic and gyrokinetic theories
 \vspace{2mm}  } \vspace{4mm}

{\large  H. Tasso$^1$, G. N. Throumoulopoulos$^2$} \vspace{2mm}

{\it $^1$Max-Planck-Institut f\"{u}r Plasmaphysik, Euratom
Association,\\
 D-85748 Garching, Germany}
\vspace{2mm}

 { \it  $^2$ University of Ioannina, Association Euratom-Hellenic Republic,\\
 Section of Theoretical Physics, GR 451 10 Ioannina, Greece }
\end{center}
%
%
\vspace{2mm}
\begin{center}
{\bf Abstract}
\end{center}
 \noindent

 A kinetic consideration of an axisymmetric equilibrium with vanishing electric field near the magnetic axis
  shows that $\bnab f$  should not vanish on axis within
 the framework of  Vlasov theory while it can either vanish or not   in the framework  of both a drift kinetic
 and
  a gyrokinetic theories ($f$ is    either the pertinent  particle or the guiding
 center distribution function).   This
 different behavior, relating to the reduction of phase space which leads to the loss of
 a Vlasov constant of motion,  may result in the  construction of different currents in the reduced phase space than the Vlasov ones.
 This conclusion is indicative of some limitation on
 the implications of reduced kinetic theories in particular as concerns  the
 physics of energetic particles in the central region of
 magnetically confined plasmas.

\newpage
 There are important phenomena in plasma physics as
 microinstabilities which can not be described in the framework of a
 macroscopic theory as magnetohydrodynamics but require the employment of kinetic theory. For the high
 temperature fusion plasmas an appropriate approximate kinetic theory  is based on the Vlasov
 equation  which  the Boltzmann/Fokker-Planck equation reduces to when the collision term is neglected. To solve self consistently the set of the
 Maxwell-Vlasov equations, however, is a tough problem related to the fact that the complete set of
 constants of motion is missing. For example, in the case of an axisymmetric equilibrium only the energy
 and the canonical momentum conjugate to the toroidal coordinate are known conserved quantities out of the four potential constants of
 motions. Because of the two missing constants of motion it is not possible to construct
  equilibria with  macroscopic   poloidal velocities, although phenomenologically sheared poloidal  velocities play an important role
  in the
 transition to improved confinement regimes of magnetically confined plasmas, e.g. the L-H transition.
 In addition, the temporary  and probably future computational
 efficiency puts a limitation  on numerical solutions of the Maxwell-Vlasov equations. For this reason approximate kinetic
 theories in a reduced phase space, as the drift kinetic \cite{Pf}-\cite{Wei} and gyrokinetic ones \cite{Ha}-\cite{WaHa}  have been developed and applied
 to numerous simulations, e.g. on turbulence in connection with the creation of  zonal flows. In
 both theories the reduced phase space is five dimensional with three spatial components associated with the guiding center
 position, $\bf R$ (or the gyrocenter position in the framework of gyrokinetic theory),  and a velocity component  parallel to the  magnetic field, $v_\paral$;
  also, the two compoments  of the
 perpendicular particle  velocity are approximated after a gyroangle averaging with the magnetic moment which is treated as an adiabatic invariant.
 A related underlying assumption  for both  reduced theories is that the ratio $\epsilon$ of the gyroperiod
 to the macroscopic time scale is small.
 In the drift kinetic theory $\epsilon$ is the same as the ratio of the gyroradius to macroscopic scale length while in
 the gyrokinetic theory small spatial variations are permitted but,
 for example,
 the amplitudes of the fluctuations to the background fields is
 equal to $\epsilon$ \cite{PfCo1}. It may be noted here that the
 reduced-phase-space kinetic theories are  developed via expansions  in $\epsilon$ the convergence of which is not
 guaranteed.

 Because of the reduction of the phase
 space some information of the particle motion is missing. This
 gives rise to the question: is the missing information important?  In the
 present note we address this question by making a comparison of
 Vlasov with  drift kinetic and gyrokinetic theories near the magnetic axis of an axisymmetric magnetically confined plasma with vanishing
 electric fields. Motivation was
 a previous study \cite{TaTh} in which by considering this equilibrium in the framework of Vlasov theory we found the following new constant of motion:
 $C_1=v_z+I \ln \left| v_\phi\right|$, where $v_\phi$ is the toroidal particle velocity, $v_z$ the  velocity component parallel to
 the axis of symmetry and $\bB_\phi = I/r \bb_\phi$  the toroidal magnetic field near axis ($r, \phi, z$ are cylindrical
 coordinates). For the sake of notation simplicity and without loss of generality here we will consider only ions
   and employ convenient units
  by setting $m=q=c=1$ where $m$ and $q$ are
 the ion mass and charge and $c$ is the velocity of light.
  Since phenomenologically the density gradients vanish on the magnetic axis it was a surprising
  conclusion of Ref. \cite{TaTh}
  that
 $\nabla f\neq 0$ must hold on axis, where $f$ is the particle distribution function. The reason is that if one assumes
 $\nabla f =0 $ thereon it is not possible   to obtain one of the known constants of motion, i.e. the
  canonical toroidal momentum $C_2=r v_\phi$.  Note also that because
 of the absolute value of $v_\phi$ in $C_1$,
 distribution functions depending on $C_1$ and the energy
 $C_3=1/2[v_\phi^2+v_z^2+v_r^2]$ can not create currents parallel to the magnetic
 field.

  The form of $C_1$ relates crucially to the toroidicity because in the straight
 case for which $\phi$ changes to $z$, $z$ to $y$ and $r$ to $x$ (where $x,y,z$
 are Cartesian coordinates) for a
 $z$-independent equilibrium with straight magnetic axis parallel
 to $z$ and arbitrary cross sectional shape  the respective constant of motion
 becomes $C_1=v_z$. Thus, in this case distribution functions depending on $C_1$ and the energy can
 produce parallel currents. It may also be noted that in the straight case  for $\bnab f \neq 0$ on axis one can obtain the
 complete set of  four constants of motion near axis: $C_1=v_z$, $C_2=v_x-B_z
 y$, $C_3=v_y+B_z x$ and $C_4=1/2(v_x^2+v_y^2+v_z^2)$ where $B_z$ is the ``toroidal" magnetic field
 on axis \cite{TaTh}.  Therefore,
 distribution functions of the form $f(C_2,C_3,C_4)$ can lead to
 purely ``poloidal" velocities near axis irrespective of the cross sectional
 shape in consistence with magnetohydrodynamics \cite{ThPa}. Unlikely,
 toroidal magnetohydrodynamic equilibria with purely poloidal velocities are not
 possible \cite{ThWe}. Whether this conclusion of nonexistence
 persists in the framework of Vlasov theory  is an open
 question relating to the two unknown constants of motion. The above
 comparison shows that the Vlasov theory well distinguishes equilibria
 with circular and straight magnetic axes.

 In the present note we  examine the same equilibrium  near axis in the
 framework of drift kinetic and gyrokinetic theories on an
 individual basis. Though the drift kinetic equations
 of Ref. \cite{CoPf} and the gyrokinetic equations of Ref. \cite{Ha} will be employed
 we claim that the conclusions do not rely on the particular forms of the reduced
 kinetic equations.
 \newline

 \noindent
 {\bf Drift kinetic theory }
 \newline

 The drift kinetic theory established in Ref. \cite{CoPf} is based on
 the Littlejohn's Lagrangian for the guiding center motion \cite{Li}   extended
 to include the polarization drift in such a way that local
 conservation of energy is guaranteed. The drift kinetic
 equation for the guiding center distribution function $f(\bR, v_\paral, \mu, t)$ (with
 $\dot{\mu}=0$)
 acquires   the form
 \beq
 \paraparb{f}{t}+ \bv\cdot\bnab f +\dot{v}_\paral
 \paraparb{f}{v_\paral}=0.
                                                   \label{1}
 \eeq
 The guiding center velocity, $\bv$, and the ``acceleration" parallel to the magnetic
 filed,
 $\dot{v}_\paral$,  can be expressed in a concise form  by introducing the modified potentials, $\bA^\star$ and
 $\Phi^\star$,
 and the respective modified electric and magnetic fields, $\bE^\star$ and
 $\bB^\star$,  as \cite{PfMo}:

  \beqa
 \bA^\star&=& \bA +v_\paral \bb +\bv_E ,  \label{2a} \\
 \Phi^\star&=& \Phi + \mu B + \frac{1}{2}(v_\paral^2+v^2_E),  \label{2b}\\
 \bv_E&=&\frac{\bE \x \bB}{B^2}, \label{2c} \\
 \bE^\star&=& -\paraparb{\bA^\star}{t}-\paraparb{\Phi^\star}{\bR}, \label{2d} \\
 \bB^\star&=&\bnab\x \bA^\star, \label{2e}
 \eeqa
 where $\Phi$ and $\bA$ are the usual electromagnetic scalar and vector  potentials and
 $\bb=\bB/B$. The quantities $\bv$ and $\dot{v}_\paral$ are then
 given by
 \beq
 \bv=\bv_g=v_\paral\frac{\bB^\star}{B_\paral^\star}+\frac{\bE^\star\x\bb}{B_\paral^\star},
                                                                 \label{3}
 \eeq
  \beq
  \dot{v}_\paral= \frac{\bE^\star\cdot\bB^\star}{B_\paral^\star}=\frac{1}{v_\paral}\bv_g\cdot\bE^\star,
                                                           \label{4}
  \eeq
  where
  \beq
  B_\paral^\star=\bB^\star\cdot \bb=B +v_\paral\bb\cdot \bnab \x\bb+ \bb\cdot
  \bnab\x\bv_E.
                                                         \label{5}
  \eeq
  Explicit expressions for $\bv$ and $\dot{v}_\paral$ are given by Eqs. (3.24) and (4.17)
  of Ref. \cite{CoPf}. Also, it is noted here that Eqs. (\ref{3})
  and (\ref{4}) have similar structure as the respective gyrokinetic
  equations of Refs. \cite{PfCo1} and \cite{CoPf1} [Eqs. (5.39) and
  (5.41) therein].
  Since $B_\paral^\star$ appears  in the denominators of (\ref{3})
  and (\ref{4}) a singularity occurs for $B_\paral^\star=0$. For $\bE=0$ this singularity can be expressed by the
  critical parallel velocity $v_c=-\Omega/(\bb\cdot \bnab \x \bb)$, where $\Omega$ is the gyrofrequency.
  Therefore,  the theory is singular for large $|v_\paral|$ at which
  $\bv$ and $\dot{v}_\paral$ diverge and consequently non-casual
  guiding center orbits occur and the guiding center conservation in
  phase space is violated \cite{CoWi}. It is the
  $v_\paral$-dependence of $\bA^\star$ [Eq. (\ref{2a})] that
  produces the singularity. In order to regularize the singularity
  $v_\paral$ in (\ref{2a}) can be replaced by an antisymmetric
  function $g(z)$ with $z=v_\paral/v_0$, where $v_0$ is some constant
  velocity \cite{CoWi}-\cite{PfMo}. The nonregularized theory presented here for simplicity
  is obtained for $g(z)=z$. In the regularized theory $g(z)\approx
  z$ should still hold for small $|z|$. For large $|z|$, however, $g$
  must stay finite such that with $v_0 \gg v_{\mbox{\small thermal}}$ one has $v_0
  g(\infty) < v_c$. A possible  choice for $g(z)$ is $g(z)=\tanh z$.

 As in the Vlasov case \cite{TaTh} we consider the drift kinetic equation
 (\ref{1}) for  an axisymmetric equilibrium  with $\bE=0$ in the vicinity of the magnetic axis. Since on
 axis the magnetic field becomes purely toroidal and dependent
 only on $r$
 one readily calculates
 $$\bnab \x \bb=\bnab \x \be_\phi=\frac{\be_z}{r},$$
 $$
 B^\star_\paral=B, \ \ \bB^\star=
 B\be_\phi+v_\paral\frac{\be_z}{r}, \ \ \bnab
 B(r)=\frac{dB}{dr}\be_r
 $$
  and consequently
 \beq
 \bv=\bv_g=v_\paral
 \be_\phi+\left(\frac{v_\paral^2}{B}-\frac{\mu}{B}\dadb{B}{R}\right)\be_z,
                                                  \label{6}
 \eeq
 \beq
  \dot{v}_\paral=0.
                                                 \label{7}
  \eeq
  As expected on axis the guiding center
 velocity consists of a component parallel to $\bB$ and the curvature
 and grad-$B$ drifts perpendicular to $\bB$  and parallel to the axis of symmetry.
 Also,  the ``acceleration" $\dot{v}_\paral$ vanishes because there is  no
 parallel force and  the drift kinetic equation (\ref{1}) becomes
\beq \bv\cdot\paraparb{f}{\bR}=0.
                                                \label{7a}
                                                  \eeq
Therefore, unlike the Vlasov description  near axis the distribution
function because of (\ref{7}) can be
 any function of $v_\paral$, which is a constant of motion, and on axis can hold either $\bnab f\neq 0$ or $\bnab f=0$.
 Thus, irrespective of the value of  $\bnab f$ on axis one  can either construct parallel currents or not by
  choosing $f$ either symmetric or antisymmetric in $v_\paral$.
 This is a significant difference from the Vlasov situation in which,  if $\bnab f =0$ on axis the
 obtainable particle distribution functions of the form $f(C_1,C_3)$ can not produce
 parallel currents.
 This discrepancy clearly relates to  missing  $C_1$ in  the
 reduced phase
 space which results in an nontrivial loss of information for the particle motion.
 Note that near axis the overwhelming majority of the particles
 are passing and the parallel currents constructed  in the framework of the drift kinetic theory
 may
 differ
 from the ``actual" ones. In the context of the drift kinetic theory the two constants of motion,
 i.e. the energy $\mu B + 1/2 v_\paral^2$
 and the canonical momentum
 $r(A_\phi+v_\paral b_\phi)$ can be found from (\ref{1}) everywhere by the method of characteristics.
 Thus, the complete set of constants of motions is obtained
 in the five dimensional phase space. These are recoverable on axis where $v_\paral$, $r$, $A_\phi(r)$ (and $\mu$)
 are conserved  individually even if
 $\bnab f=0$.  Unlikely, in the respective Vlasov case  the toroidal angular momentum constant of motion, $C_2$,
  is missed  when $\bnab f=0$ is
 assumed.  Also, for straight $z$-independent equilibria one has on  axis $\bnab \x \bb=\bnab \x
 \be_z=0$, $\bB^\star=\bB \be_z$, $\bv_g=v_\paral \be_z$,
 $\dot{v}_\paral=0$ and (\ref{1}) is identically satisfied, implying
 that $f$ can be any function of $x,y, v_\paral$ and $ \mu$.
 Therefore, unlike the Vlasov theory the dependence of $f$ on $v_\paral$ is independent of
 toroidicity and therefore,
 regarding  the formation of parallel currents, the drift kinetic theory  can
 not distinguish equilibria of circular or straight magnetic axes.
 \newline

 \noindent
 {\bf Gyrokinetic theory }
 \newline

 We will use the gyrokinetic equations of Ref. \cite{Ha} which
 have been employed to a variety of applications (see for example  the recent Refs.
 \cite{LaMc,FeLi,ZhLi}). Eq. (\ref{1}) remains identical in form
 where
 $f(\bR, v_\parallel, \mu, t)$ is now the gyrocenter distribution function
 for ions. The gyrocenter velocity and ``acceleration" are given by
 \beq
 \bv=\bv_g=v_\paral \bb_0
 +\frac{B_0}{B_{0\paral}^\star}\left(\bv_E+\bv_{\nabla
 B}+\bv_c\right),
                                                    \label{8}
 \eeq
 \beq
 \dot{v}_\paral=-\frac{1}{v_\paral}\bv_g\cdot\left(\bnab
 \overline{\Phi}+\mu \bnab B_0\right).
                                                    \label{9}
 \eeq
 Here, $\bB_0$ is the equilibrium magnetic field, $\bb_0=\bB_0/B_0$,
 $$B_{0\paral}^\star=(\bB_0+v_\paral\bnab\x\bb_0)\cdot \bb_0,$$
 $ \overline{\Phi}$ stands for the perturbed gyroaveraged
 electrostatic potential, and the  $\bE\x\bB$-drift velocity $\bv_E$, the grad-$B$ drift velocity
 $\bv_{\nabla B}$,  and the curvature drift velocity $\bv_c$ are given
 by
 \beq
 \bv_E=-\frac{\bnab \overline{\Phi}\x\bnab B_0}{B_0^2},
                                                    \label{10}
 \eeq
 \beq
 \bv_{\nabla B}=\frac{\mu}{\Omega B_0}\bB_0\x\bnab B_0,
                                                   \label{11}
 \eeq
 \beq
 \bv_c=\frac{\mu v_\paral^2}{\Omega B_0^2}
 \bb_0\x\bnab\left(p_0+\frac{B_0^2}{2}\right).
                                                \label{12}
 \eeq
 Note that as in the case of  drift kinetic theory a similar singularity occurs at
 $B_{0\paral}^\star=0$. In numerical applications this singularity was
 ``avoided"    by approximating $B_{0\paral}^\star=B_0$ (see for example  Refs.   \cite{LaMc,FeLi}).
 Consideration of the above equations  for an axisymmetric equilibrium with $\bE=0$ near axis
 yields  relations similar  to (\ref{6}), (\ref{7}) and (\ref{7a}). Therefore,  the  above
 found
  discrepancies of the drift kinetic theory
  with the Vlasov one persist in the framework of the gyrokinetic theory.
  The structure of the reduced kinetic equations in conjunction with
  the symmetry of the equilibrium considered  clearly indicate that
  this conclusion is independent of the particular drift kinetic or
  gyrokinetic equations.

  In conclusion, first a singularity
  which occurs in both drift kinetic and gyrokinetic theories  for large parallel
  particle velocities is usually eliminated in the literature by a rough
  approximation. Second, a  comparison of the Vlasov equation with either the
  drift kinetic or the gyrokinetic equation near the magnetic axis
  of an axisymmetric equilibrium with vanishing electric field
  implies different properties of $\bnab f$   and, unlike Vlasov,  non distinguishing of  equilibria  with
  straight and circular magnetic axes in connection with the formation of parallel currents.
   This relates to  the loss of
  a Vlasov constant of motion in the reduced phase space. Consequently,  different drift kinetic or gyrokinetic parallel currents may be
  created than the Vlasov ones.
  This indicates that the reduction of the
  phase space,  even  if  made rigorously so that local conservation laws and Liouvillean
  invariance of the volume element
   is guarantied, is associated with the loss of  nontrivial physics  which could put certain limits on the validity
   of the conclusions from drift kinetic or gyrokinetic
   simulations.

 \section*{Acknowledgements}


 Part of this work was conducted during a visit of  G.N.T.
 to the Max-Planck-Institut f\"{u}r Plasmaphysik, Garching.
 The hospitality of that Institute is greatly appreciated.

 This work was performed  within the participation of the University
of Ioannina in the Association Euratom-Hellenic Republic, which is
supported in part by the European Union and by the General
Secretariat of Research and Technology of Greece. The views and
opinions expressed herein do not necessarily reflect those of the
European Commission.
\newpage

\vspace*{-1.5cm}

 \end{document}